%% file: acl_latex.tex
\pdfoutput=1

\documentclass[11pt]{article}

\usepackage[preprint]{acl}

\usepackage{times}
\usepackage{latexsym}

\usepackage[T1]{fontenc}

\usepackage[utf8]{inputenc}

\usepackage{microtype}

\usepackage{inconsolata}
\usepackage{graphicx}
\usepackage{float}
\usepackage{subfigure}
\usepackage{booktabs} %
\usepackage{multirow}
\usepackage{hyperref}

\usepackage{amsmath}

\title{Efficient Title Reranker for Fast and Improved Knowledge-Intense NLP}

\author{Ziyi Chen, 
 {\bf Jize Jiang}, 
 {\bf Daqian Zuo}, 
 {\bf Heyi Tao},
 {\bf Jun Yang},\\ 
 {\bf Yuxiang Wei},
 {\bf Kevin Chang} \\
  University of Illinois Urbana-Champaign, IL, USA \\
  \texttt{\{ziyic2, kcchang\}@illinois.edu} \\}

\begin{document}
\maketitle

\input{sec/0_abstract}
\input{sec/1_intro}
\input{sec/2_background}

\input{sec/3_method}

\input{sec/4_loss}

\input{sec/5_experiments}

\input{sec/6_related_work}

\input{sec/7_conclusions}
\input{sec/8_limitations}

\nocite{CorpusBrain, MultiDPR, SEAL, hofstätter2022fidlight, bai2023griprank} 
\bibliography{custom}

\appendix

\end{document}

%% file: sec/0_abstract.tex
\begin{abstract}

In recent RAG approaches, rerankers play a pivotal role in refining retrieval accuracy with the ability of revealing logical relations for each pair of query and text.
However, existing rerankers are required to repeatedly encode the query and a large number of long retrieved text. This results in high computational cost and limits the number of retrieved text, hindering accuracy.
As a remedy for the problem, we introduce the Efficient Title Reranker via Broadcasting Query Encoder, a novel technique for title reranking that achieves a 20x-40x speedup over the vanilla passage reranker.
Furthermore, we introduce Sigmoid Trick, a novel loss function customized for title reranking. 
Combining both techniques, we empirically validated their effectiveness, achieving state-of-the-art results on all four datasets we experimented with from the KILT knowledge benchmark.

\end{abstract}

%% file: sec/1_intro.tex
\section{Introduction}

Recently, many large language models have shown promising abilities to generate texts in fluent English and are able to perform a wide range of natural language processing tasks. \cite{openai2023gpt4, chowdhery2022palm, brown2020language} However, large language models suffer from issues such as hallucination \cite{zhang2023siren} and the difficulty in keeping parameters update-to-date which hinders the practical use of large language models. 
One of the methods that can alleviate these issues is Retrieval-Augmented Generation (RAG) \cite{lewis2021retrievalaugmented, wang2023survey}. By leveraging retrieved information from a trusted corpus, retrieval context can help reduce hallucination and let the model retrieve up-to-date information in the context \cite{lewis2020retrieval}.

\begin{figure}[t]
\centering
\includegraphics[width=0.49\textwidth]{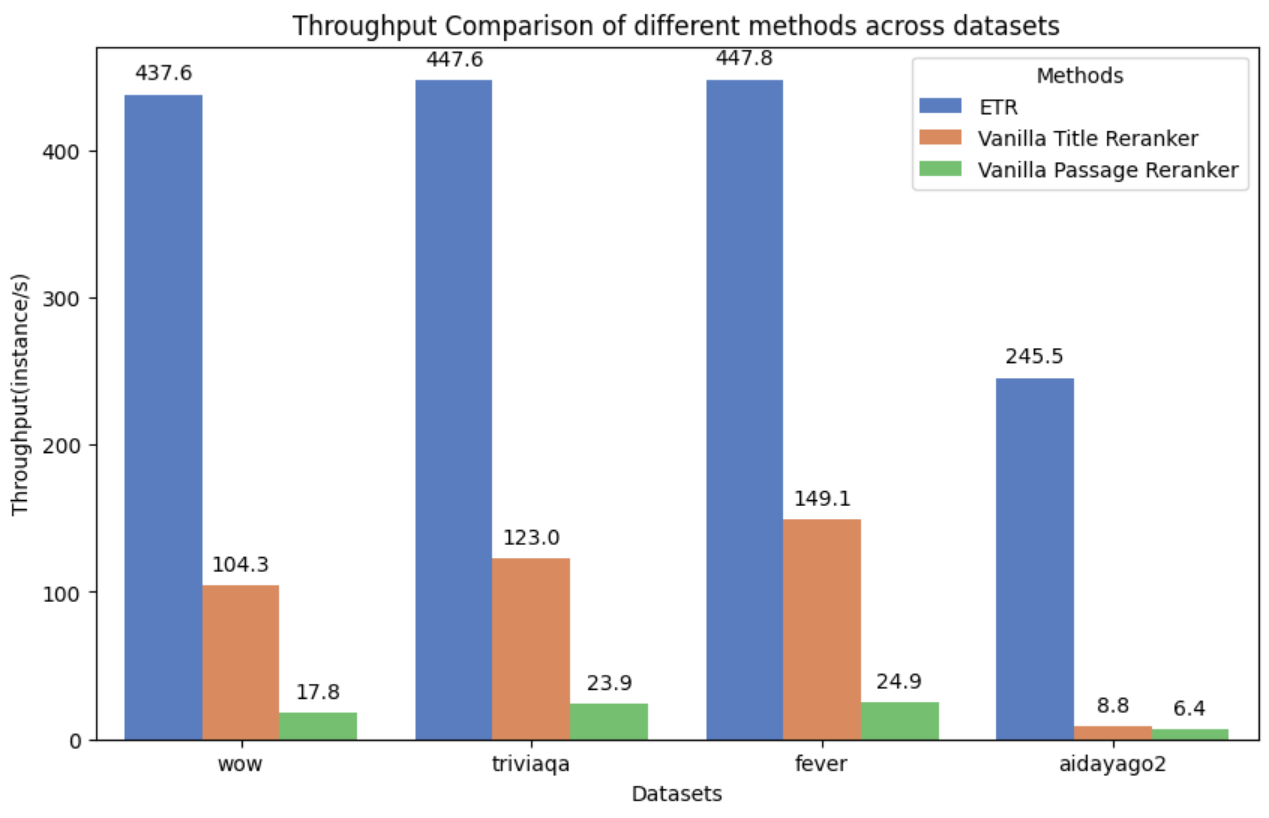}\hfill
\caption{ A comparison of throughput of different methods on a single Nvidia A40.
Our method (ETR) via Broadcasting Query Encoder (BQE) has a throughput improvement of 20x to 40x compared to vanilla passage rerankers, and at least 3x throughput improvement over normal title reranker. \textit{Best viewed in color.}}
\label{fig:speedup}
\end{figure}

Recent RAG approaches typically include two stages of retrieval pipeline. Sparse or dense retrieval as the first stage, and a reranker as a second stage. 
\cite{glass2022re2g, hofstätter2022fidlight, bai2023griprank}
Dense retrieval leverages the distance between the embeddings of the query and the passage serves as the score \cite{lewis2020retrieval}. This results in enhanced efficiency; however, it lacks the expressiveness to handle many-to-many relations. 
In comparison, a reranker scores each query and text pair individually, making it possible to express logical relation between each pair of query and text.
Consequently, rerankers are leveraged for the expressibility in recent State-of-the-Art RAG systems \cite{nogueira2020passage, nogueira2019multistage, glass2022re2g}.

However, existing approaches use documents or long passages for retrieval, so the number of candidates that can be processed in one run of the model is limited due to memory constrain, making many users unable to fully utilize the potential of rerankers due to computational constraints.  

This reveals a new approach of reranking by using condensed textual representation such as titles, URLs, or summaries for reranking. Since condense textual representation is much shorter than documents or passages existing reranker reranks, so a large number of condense textual representation can be reranked for better utilization of expressibility of rerankers.

Reranking of condensed textual representation is different from existing methods and poses unique challenges. 
Existing exploration of rerankers in RAG systems are mainly focused on passages of at least 100 words \cite{glass2022re2g, hofstätter2022fidlight, bai2023griprank}. 
Therefore, simple adoption of these approach might not maximize the efficiency when reranking condensed textual representation; in addition,  
the condense textual representation is much shorter. Without careful construction, condensed textual representation can be void of vital information, making training of a reranker a challenge. 
In this paper, our focus is the efficiency as well as the efficacy for reranking of condensed textual representation.

In our work,
we introduce \textbf{Efficient Title Reranker (ETR)} using a novel technique \textbf{Broadcasting Query Encoder (BQE)} to drastically reduce the reranking time for retrieval tasks.
Specifically, ETR involves reranking only the condensed textual representation of the retrieved passages. This greatly reduces the number of tokens to be reranked; in addition, the BQE technique achieves further speedup by avoiding repeated encoding of potentially long queries. This is achieved by manipulating the attention masks and position bias, as shown in Figure \ref{fig:BQE}. 
Empirically, we found the technique of BQE success as ETR achieving 20x to 40x speedup compared to vanilla passage reranker as shown in Figure \ref{fig:speedup}.

Despite the architectural association between ETR and passage reranking, there are significant differences between condensed textual representation reranking and passage reranking. Due to the limited length of a title, it can be void of vital information relating to the query, or it can be part of the query making it an easy answer. 
These properties are undesirable for training and can lead to reduced reranking accuracy. 

In search of a solution, we notice an association between the sigmoid function and title reranking - training example with either property having probability of extreme values, corresponding to the region where the sigmoid function plateaus. 
This makes the sigmoid function an ideal candidate for reducing gradient updates on these training examples.
In light of this, we introduce
\textbf{Sigmoid Trick} that penalizes both extremities to improve training process. 
The sigmoid trick involves using an adjusted sigmoid function as a wrapper such that the aforementioned problematic training samples will receive reduced gradients, leading to better efficacy in training.

To validate the effectiveness of our approach, we use the Knowledge-Intensive Language Tasks (KILT) benchmark~\cite{petroni2021kilt} for evaluation. 
The KILT benchmark contains a diversified collection of tasks, and it was introduced to evaluate the capabilities of pre-training language models to address NLP tasks that require access to external knowledge. We evaluated our approach on a range of tasks on corresponding datasets including question answering on TriviaQA~\cite{joshi2017triviaqa}, entity linking on AIDA CoNLL-YAGO~\cite{decao2020autoregressive}, dialog generation on Wizard of Wikipedia~\cite{dinan2018wizard}, and fact checking on FEVER~\cite{thorne2018fever}. During our experiments on these benchmarks, ETR has achieved SotA on all benchmarks in recall@5 metric, showing the effectiveness of ETR over existing methods.\\
Our main contributions are:
\vspace{-5pt}
\begin{itemize}
\setlength\itemsep{0em}
\item We introduced the novel Broadcasting Query Encoder technique to implement an Efficient Title Reranker that enables a one-pass fast reranking of titles. Improving the reranking speed by 20x to 40x times over the vanilla passage reranker.
\item We proposed the Sigmoid Trick, a method of loss function modification that improves training of ETR.
\item Experimented showed the effectiveness of the Sigmoid Trick and Efficient Title Reranker as they combined achieved multiple SotA on kilt knowledge benchmark.

\end{itemize}

\vspace{-10pt}

\begin{figure*}
    \centering
    \includegraphics[width=1.0\textwidth]{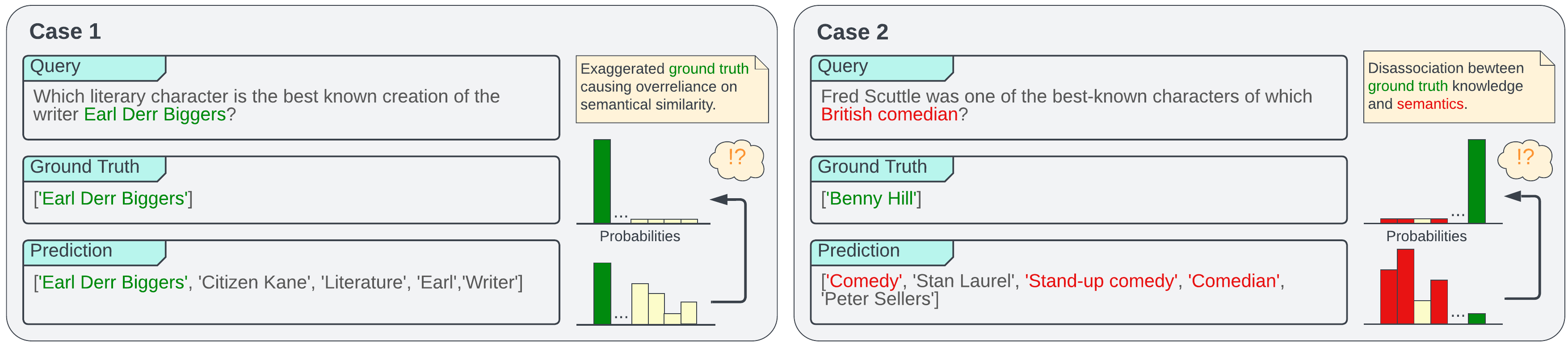}
    \caption{
    Illustration of the two undesirable training scenarios. 
    In the first case,'Eearl Derr Biggers' is been correctly predicted, but this example is trivial. Overtraining on this example will make the model only use semantical similarity and fail to handle more difficult queries. 
    In the second case, the prediction failed to include 'Benny Hill', the ground truth label.
    However, since there is no information from the query for the model to judge whether the ground truth is correct. Examples like such become a "noise" during the training process. 
    \textit{Best viewed in color.}
    }
    \label{fig:intro_example}
\end{figure*}

%% file: sec/2_background.tex
\section{Preliminary}
Informational retrieval usually involves an indexed search such as BM25 or Dense retrieval \cite{karpukhin2020dense} that retrieves related documents from a large corpus. Despite recent advancements \cite{ren-etal-2021-rocketqav2, 10.1145/3543507.3583421}, indexed searches still lack the expressibility to score individually for a text-query pair, so a reranker becomes a significant component in recent information retrieval systems.
We review existing methods of neural rerankers that serve as the foundation of our study.

\textbf{Text reranking} 
For a query $q$ and textual information $t_1, t_2,..,t_n$, the task of text reranking is to generate a series of scores $s_1, s_2,..., s_n$ each corresponding to the input textual information. The score $s_i$ represents the association between the query $q$ and textual information $t_i$. Such association is specified in each downstream task, e.g. evidence to support or refute in FEVER \cite{thorne2018fever}, the related Wikipedia article of the conversation in Wizard of Wikipedia \cite{dinan2018wizard}.\looseness=-1

\textbf{Pretrained language models for neural rerankers} 
Pretrained language models are leveraged for text reranking in many previous works due to their superior textual understanding ability. 
A pretrained language model can be fine-tuned to be a textual reranker, and large language models can also be used zero-shot. 
\citet{nogueira2020passage} first attempted to re-purpose BERT \cite{devlin-etal-2019-bert} to be a reranker by fine-tuning it on the MS MARCO dataset. The approach is named monoBERT.
Later, a similar approach was used on T5 \cite{t5_original} resulting in monoT5 model \cite{nogueira2020document}. 
Recently, with the popularity of large language models such as GPT4 \cite{openai2023gpt4} in zero shot, attempts have been made to use APIs of ChatGPT and GPT4 to perform reranking \cite{sun-etal-2023-chatgpt}.  \looseness=-1

\textbf{Design of a monoReranker} 
Both monoBERT \cite{nogueira2020passage} and monoT5 \cite{nogueira2020document} are considered as monoRerankers.
Scoring of a monoReranker is a binary classification task over individual candidates. To generate a score, either existing model weight or a new project layer is used to map the last latent state into two logits \cite{nogueira2020passage, nogueira2020document}; the score, being the output probability achieved via softmax on the logits, is used for reranking. 
In the design of monoBERT, the two-dimensional logits are achieved by mapping the vector of [CLS] into a two-dimensional vector. For monoT5, the mapping is taken from two columns of the language modeling head corresponding to the "true" token and the "false" token.
The logits are then transformed into ranking scores via softmax function.
\looseness=-1

%% file: sec/3_method.tex
\begin{figure*}[ht]
    \centering
    \includegraphics[width=1.0\textwidth]{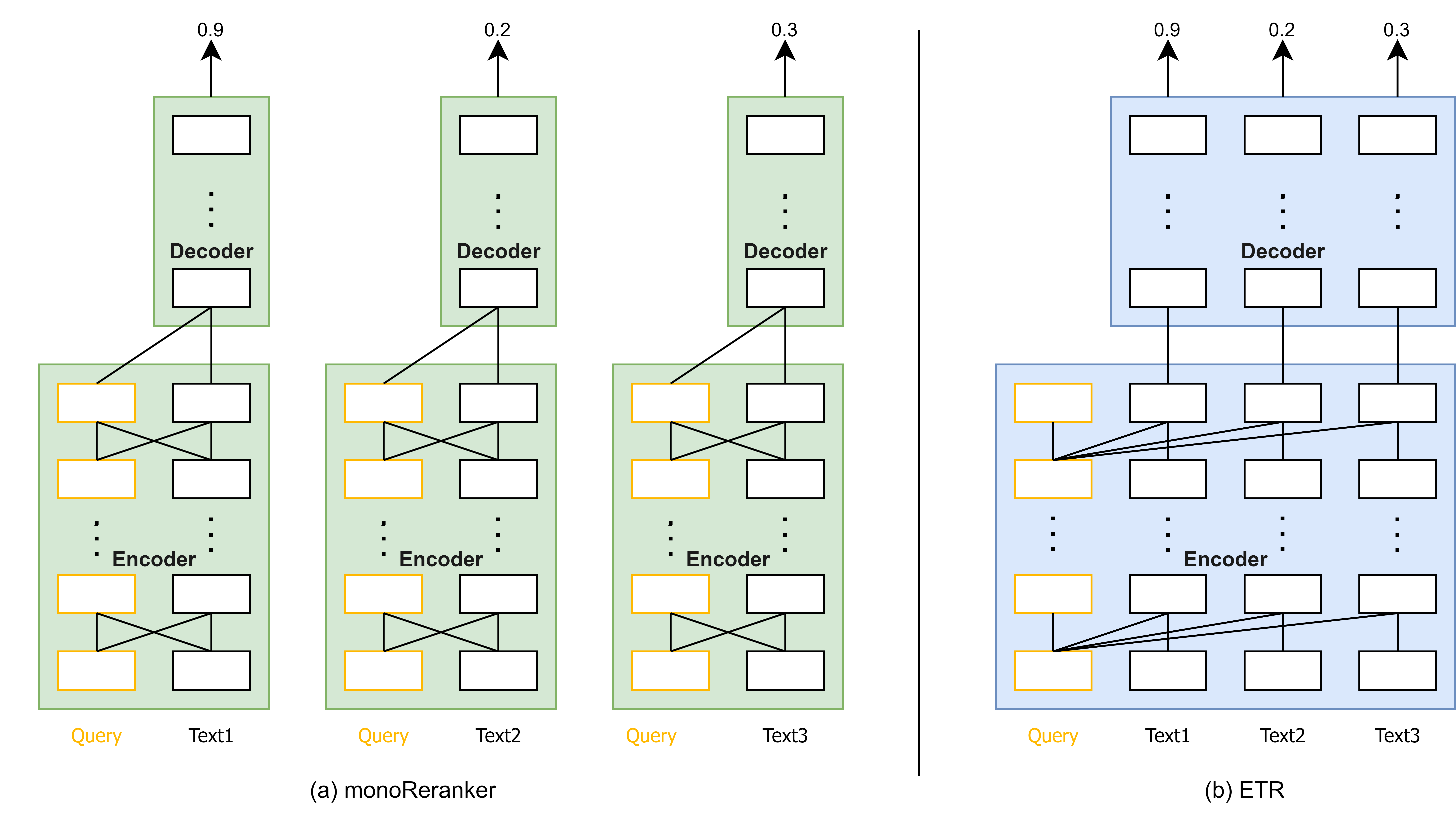}
    \caption{A comparison between monoReranker and ETR. ETR only encoded the query once to score for multiple texts while monoReranker will need to encode the query as many times as the number of the text.
    The reduced number of encoding of the query in ETR is achieved via attention manipulation that alloww each title to be scored individually while being encoded in one model run. }
    \label{fig:BQE}
\end{figure*}

\begin{table}[h]
  \centering
  \begin{tabular}{ | c |  c | } 
    \hline
    \bfseries dataset input & \bfseries Token count   \\
    
    \hline
    Aidayago2     &  624.47 \\
    FEVER     &  13.88 \\
    WOW     &  93.63 \\
    TriviaQA     &  21.25 \\
    
    \hline
    \bfseries Article Title & \bfseries Token count   \\
    
    \hline
    Wikipedia Title     &  4.0 \\
    \hline
  \end{tabular}
  \caption{Average number of tokens in the input query of our experimented datasets compared to the average number of tokens in the Wikipedia titles. Tokenization is done using T5-tokenizer.}
  \label{table:token_count}
\end{table}

\section{Efficient Title Reranker via Broadcasting Query Encoder}

In retrieval-augmented generation tasks, the recall of the entire system is upper-bounded by the recall of passages forward to the reranking stage. Therefore, given a powerful reranker, increasing the number of reranked passages can potentially improve the recall of the system. 
For example, using BM25 alone on the FEVER dev set in KILT knowledge benchmark will result in a recall@30 of 82.12 and recall@100 of 88.41, yielding an improvement of 6.29 on the upper bound by increasing the number of passages from 30 to 100. 
Yet, an increase in the number of reranked passages also costs additional computational resources. This is undesirable in practice and can be challenging for large-scale services. Therefore, improving the efficiency of reranker becomes a key problem.

A key limiting factor of reranker efficiency resides in the large number of input tokens. To rerank 100 passages, a rerank needs to encode more than ten thousands tokens. However, the input token can be largely reduced by leveraging condensed textual representation that succinctly represent the content of the documents and are sufficient for many queries. 
Inspired by the observation,
we propose \textit{title reranker} - a novel reranker that only reranks titles. With an average length title on Wikipedia is only 4 tokens, the input of title reranking is drastically shortened compared to the passages resulting in a significant speedup.\looseness=-1

Despite the speedup of title rerankers, there is still an inefficiency remaining - the repeated encoding of queries. 
As shown in table \ref{table:token_count}, the input queries are on average much longer than the titles across different datasets.
Among the four datasets the longest one, Aidayago2, has an average queries length of 156.12 times the average Wikipedia title.  
Since the reranker needs to encode the query with each of the text to be reranked, repeated encoding of long queries consume the majority of the memory, becoming the limiting factor of the throughput.\looseness=-1

To resolve the dilemma, we propose \textbf{E}fficient \textbf{T}itle \textbf{R}eranker (ETR) via \textbf{B}roadcasting \textbf{Q}uery \textbf{E}ncoder (BQE) which only encodes the query once for reranking of multiple titles. The reduced encoding of the query is achieved by manipulating attention and position bias. 
In a run of the model, we feed both the query and the titles into the encoder, and to avoid titles attending to each other or the query attending to multiple titles, we adjust the attention mask such that the query can only attend to itself while titles can attend to both itself and the query, as shown in Figure \ref{fig:BQE}. 
In addition, the original positional embedding will make a variance to titles at different positions which is undesirable, so we revised the positional embedding of the title to be immediately after the query.
From there, we create a list of initial tokens as many as the titles for decoder input, and we limit each initial token to only be able to attend to the corresponding title to score. To generate the score, we used the monoT5 design introduced by \citet{nogueira2020document}; we harvest only the score from \textit{Yes} and \textit{No} and take softmax over these two logits to achieve a probability to rank the titles. Overall, from the perspective of the model, the design of BQE makes the reranking process of ETR being very close to the vanilla mono-reranker except for not allowing the query to attend to the titles.

%% file: sec/4_loss.tex
\section{Sigmoid Trick} \label{Sigmoid Trick}

Because titles are much more condensed in information, training of title reranking faces different challenges compared to passage reranking training. Most training datasets available for retrieval tasks contain human annotations for passage reranking~\cite{petroni2021kilt} meaning the ground truth passages contain supportive information for the query. 
The same guarantee doesn't always hold for title reranking as the title can have little to no association with the query. 

As shown in figure \ref{fig:intro_example}, the example on the right represents the disassociation between the query and the ground truth, making it a noise during training without prior knowledge.
On the other hand, some training is too easy since the ground truth is part of the query. Training on these easy answers can cause the model to over-relying on semantical similarity. 
These two types of problematic training examples of title reranking pose challenges to training a title reranker, making a tailored method essential for training a title reranker.

\begin{figure}[t]
  \centering
  \includegraphics[width=1.0\linewidth]{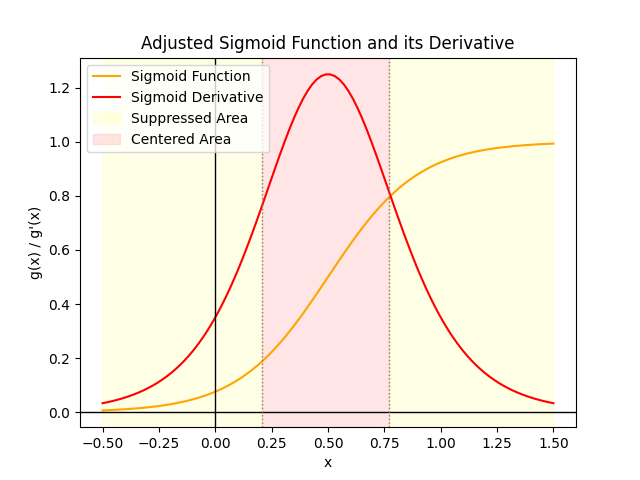}
  \caption{
  A plot of $y = S(5 (x - 0.5))$ and its derivative where $S$ being the sigmoid funtion. The derivative of the sigmoid function will decrease as it gets further away from the center. The red and yellow region shows the area that receives at least 0.8x of gradient update and at most 0.8x of gradient update.}
  \label{fig:sigmoid_trick}
\end{figure}

In attempts to address the issue, 
we found an association between the sigmoid function and the probability of these two problematic training examples. 
The derivative of the sigmoid function $S$ has its highest value at its center and diminishes as its input derivates from the center as shown in Figure \ref{fig:sigmoid_trick}. 
On the other hand, both problematic training examples have a probability away from the center of the probability interval, 0.5. For the training samples with easy answers, the probability is close to 1 due to its strong semantical similarity; on the other hand, noisy training samples will have probabilities close to 0. 
Therefore, the sigmoid function has a desirable property that can be used to suppress the gradient of both training examples by setting the center to 0.5. 

To combine the sigmoid and the loss function to suppress the two problematic training examples, we use a basic formula in calculus, the chain rule. Given the model parameter $\theta$, original loss function $\mathcal{L}$, sigmoid function $S$, the following equation follows from the chain rule,
\begin{equation*}
    \frac{dS(\mathcal{L}(\theta))}{d\theta} =  S'(\mathcal{L}(\theta))\frac{d\mathcal{L}(\theta)}{d\theta} 
\end{equation*}
In backpropagation, the term $S'(\mathcal{L}(\theta))$ will serve as a weight to the graduate. As the loss function is away from the center, the term $S'(\mathcal{L}(\theta))$ will diminish to suppress the gradient of the problematic training samples.

Following the intuitive interpretation above, we provide a few different variations of the sigmoid trick.

Given a hyperparameter $\epsilon$ adjusting the penalty of sigmoid trick. Let $y^{+}, y_1^{-},...,y_k^-$ be the probabilities of positive sample and negative samples, the \textit{sigmoid contrastive loss} is defined as the following 
\begin{equation*}
\begin{split}
    \mathcal{L}_{sig\_con}&(y^{+}, y_1^{-},...,y_k^-) =\\
    &- S(\epsilon (\frac{y^{+}}{y^{+} + \frac{1}{k}\sum_{i=1}^{k} y_i^{-}} - 0.5))
\end{split}
\end{equation*}
The sigmoid contrastive loss penalizes when the probability of the positive sample deviates from the average of negative sample probabilities.
However, sigmoid contrastive loss has its own issue.
When the average of negative samples is near 0, the sigmoid contrastive loss can reduce the gradient flow even when the probability of ground truth is not near 0 or 1 resulting in a decrease in reranking accuracy.

Therefore, as a solution to the above issue, we propose a different usage of the sigmoid trick that treats the positive and negative samples separately. \looseness=-1
With a hyperparameter $\lambda_{gt}$ and $\lambda_{neg}$ representing a preset average of positive and negative examples. The separated sigmoid loss is the following:
\begin{equation*}
\begin{split}
    \mathcal{L}_{sep\_sig}&(y^{+}, y_1^{-},...,y_k^-) =\\
    &- S(\epsilon (y^{+} - \lambda_{gt})) \\
    &- S(\epsilon (\lambda_{neg}- \frac{1}{k}\sum_{i=1}^{k} y_i^{-}))
\end{split}
\end{equation*}
The separation of ground truth probability and negative probability means the issue of sigmoid contrastive loss will not affect separated sigmoid loss. 
Yet, there is also a limitation with separated sigmoid loss. Since the average of ground truth probability and the negative samples' probabilities can change over time, having a fix $\lambda_{gt}$ and $\lambda_{neg}$ might limit the training when the average of ground truth probability or negative samples' probabilities deviate from $\lambda_{gt}$ and $\lambda_{neg}$ respectively; noticeably, for this case, sigmoid contrastive loss are less affected due to the dynamic property of contrastive loss.

Although both sigmoid contrastive loss and separated sigmoid loss have their issue, their issues are mostly disjoint. Consequently, a combination of these two losses can alleviate both issues leading to higher performance.
The combined sigmoid loss is defined as the following:
\begin{equation*}
\begin{split}
    \mathcal{L}_{combined\_sig}&(y^{+}, y_1^{-},...,y_k^-) = \\
    \gamma & \mathcal{L}_{sig\_con}(y^{+}, y_1^{-},...,y_k^-) + \\
   (1- \gamma)  & \mathcal{L}_{sep\_sig}(y^{+}, y_1^{-},...,y_k^-) 
\end{split}
\end{equation*}

%% file: sec/5_experiments.tex
\section{Experiments}

\begin{table*}
  \centering
  \begin{tabular}{lllllllll}
    \hline
    \bfseries Architecture & 
    \multicolumn{1}{c}{\bfseries Fever} & \multicolumn{1}{c}{\bfseries Wow} & \multicolumn{1}{c}{\bfseries TriviaQA} & \multicolumn{1}{c}{\bfseries Aidayago2} \\
    \cline{2-3}  \cline{4-5} \cline{6-7} \cline{8-9} 
    
    & Recall@5 & Recall@5 & Recall@5 & Recall@5 \\ 
    \hline
    RAG     & 75.55 & 74.61 & 57.13 & 72.62 \\
    SEAL     & 89.56 & 78.96 & 76.36 & - \\
    MultiDPR   & 87.52 & 67.13 & 68.33 & 	39.46 \\
    GENRE     & 88.15 & 77.74 & 75.07 & 94.76 \\
    CorpusBrain    & 90.50 & 	\underline{81.85} & 75.64 & \underline{94.85} \\
    Re$^2$G    & \underline{92.52}  &  79.98  & 74.23 & - \\
    FiD-Light &  92.26   & 76.51 & 78.54 & - \\
    GripBank &  -   & 76.51 & \underline{78.83} & - \\
     ETR(our)   & \textbf{92.66} & \textbf{83.75} & \textbf{85.00} & \textbf{96.44}\\
    \hline
  \end{tabular}
  \caption{The evaluation result on KILT benchmark on test set hosted by EvalAI. We use recall@5 as our evaluation metric for retrieval results. The best result for each dataset is bolded. The second-best is underlined. }
  \label{table:result_table}
\end{table*}

\subsection{Datasets} 
We evaluate ETR on four tasks in the KILT knowledge benchmark \cite{petroni2021kilt}. 

\textbf{Open-Domain Question Answering} we use TriviaQA \cite{joshi2017triviaqa} to evaluate the ability of our models in the task of open-domain question answering that requires the model to retrieve the correct evidence based on the input question. 

\textbf{Knowledge-Enhanced Dialogue Generation}
    The evaluation for knowledge-enhanced dialogue generation is conducted using the Wizard of Wikipedia (Wow) dataset \cite{dinan2018wizard}. In this task, the input comprises a dialogue history culminating in the information seeker's turn, and the model is expected to ground the correct Wikipedia page for a specific knowledge sentence.
    
\textbf{Fact-Checking} For fact-checking, the Fact Extraction and VERification (Fever) dataset \cite{thorne2018fever} is employed. The model needs to retrieve Wikipedia articles as evidence to support or refute based on the input claim.

\textbf{Entity Linking}
    In our evaluation of Entity Linking, we focus on the AIDA CoNLL-YAGO (Aidayago2) dataset. The datasets contain a long paragraph with a single entity tagged with special tokens, namely [START\_ENT] and [END\_ENT]. The goal is to output the title of the Wikipedia page corresponding to the tagged entity.

\subsection{Knowledge-intensive Natural Language Processing}

\textbf{Model Selection}
Our ETR models are based on FLAN-T5 models \cite{chung2022scaling, t5_original}. Specifically, to get the best performance possible, we choose to use FLAN-T5-xl, with 3 billion parameters, the largest model we can train with our NVIDIA A40 GPUs. 

\textbf{Implementation}
We implemented the broadcasting query encoder via manipulation of attention and position bias. We adjust the encoder attention mask such that attention exists among tokens of each title and the query, and from titles to the query. To eliminate the effect of position bias, we adjust the position bias such that the position of each title is immediately after the query.

\textbf{Initial Retrieval} We use both GENRE \cite{decao2021autoregressive} and BM25 to perform the initial retrieval to be passed to the reranker models. We retrieved 5 using GENRE and 995 using BM25 for each datasets except for aidayago2. We retrieve 50 using GENRE and 1000 from BM25 for Aidayago since BM25 doesn't perform well.
During training, we used 5 GENRE retrieved titles and 35 randomly sampled BM25 passages as negatives.

\textbf{Evaluation}  During evaluation, we merged the GENRE result and BM25 result equally and took the first 10-400 passages for reranking to decide the best number of passages for each datasets.
We then run on test set and submit our results to 
EvalAI\footnote{\url{https://eval.ai/web/challenges/challenge-page/689/leaderboard}} for evaluation.
With the advancement of sparse attention \cite{child2019generating} and new architecture supporting linear complexity scaling with respect to input length \cite{fu2023hungry, gu2023mamba}, multiple retrieved articles can be forward to LLMs for higher performance; therefore, we focus on recall@5 instead of R-Precision as the evaluation metric.\looseness=-1

\textbf{Baseline} We choose the best-performing methods existing on KILT Knowledge benchmark as well as some popular methods as our baseline.

\textbf{Result} As shown in Table \ref{table:result_table}, in all four datasets we experimented with, we achieved state-of-the-art retrieval performance measured by recall@5. Specually, on TriviaQA, Wizard of Wikipedia, and Aidayago2, our framework surpasses current SotA by a large margin. The result validates the novel concept of title reranking as an effective way for knowledge-intensive natural language processing tasks.\looseness=-1

\subsection{Efficiency vs Accuracy}

\begin{figure}[h]
    \centering
    \includegraphics[width=\linewidth]{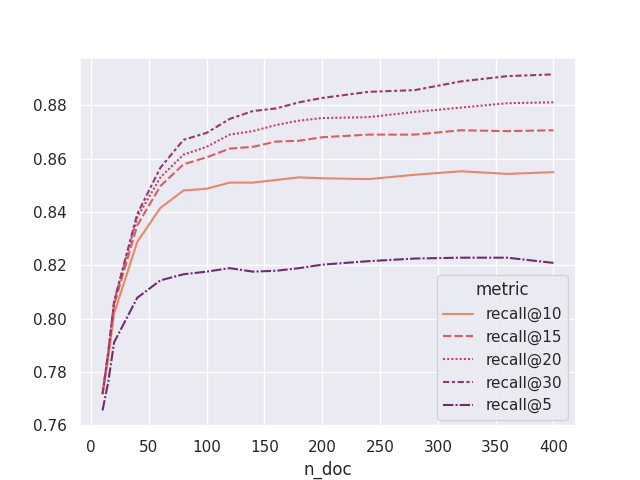}
    \caption{The recall for ETR on Wizard of Wikipedia test by the number of retrieved documents.}
    \label{fig:n_doc}
\end{figure}

To understand ideal number of retrieved documents for ETR to rerank, we evaluated our reranker with different numbers of reranking documents ranging from 10 to 400 as shown in \ref{fig:n_doc}. 
All recall percentages increase as the number of documents reranked increases up to 300. Some recall percentage still increase as the number of document increase to 400.

The result suggests that increasing the number of titles for ETR does lead to an improvement in the reranking efficacy measured by recall score at different numbers of passages. This also suggests that with a limited computational budget, improving the efficiency of reranker, which allows for more titles to be reranked, can lead to a better improvement in reranking efficacy.

\vspace{-1.0pt}
\subsection{Hyperparameter of Sigmoid Trick}
To better understand the effect of the hyperparameter $\epsilon$ in sigmoid trick, we performed experiments to reveal the relationship between the value of $\epsilon$ and effectiveness of training.

\textbf{Setup} Since the experiment requires large number of runs, we use FLAN-T5-small for its efficiency. We experiment with $\epsilon$ from 1 to 21 with an increment of 4. We split the dev set into two equal part serving as validation and test set.
We fine-tune on each of the dataset for up to 40,000 training samples, and evaluation on validation half of the dev set every 10,000 steps. 
We evaluate by comparing the probability of ground truths and top 100 BM25 negative samples to decide the rank of the ground truth as a measure avoid dependency on the first stage retrieval results.
We choose the best checkpoint on validation half of the dev set to be evaluated on the test half of the dev set. 

\begin{figure*}[t]
  \centering
  \includegraphics[width=1.0\linewidth]{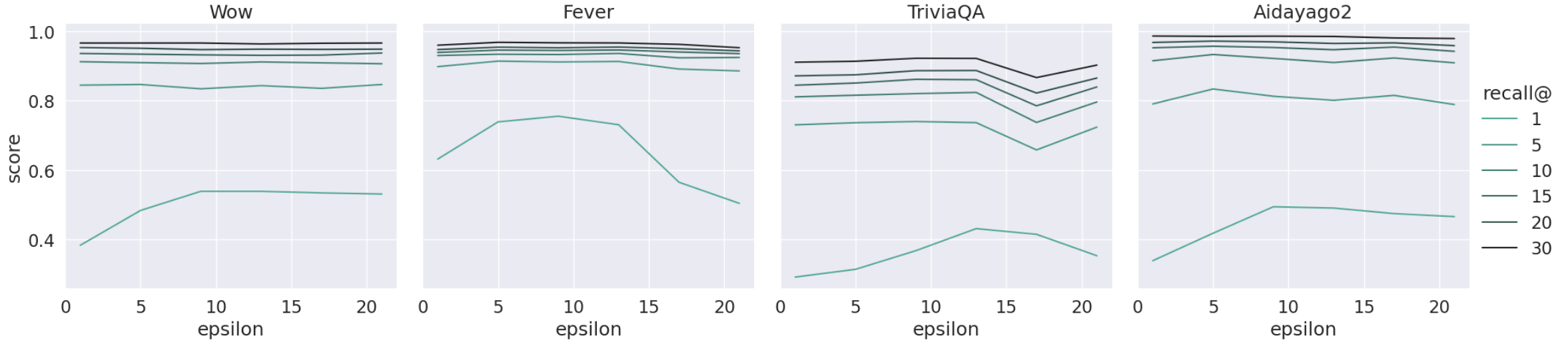}
  \caption{
  Plots of different retrieval metrics against epsilon, the hyperparameter of sigmoid loss, across four datasets. }
  \label{fig:sigmoid_comparison}
\end{figure*}

\textbf{Result} As shown in Figure \ref{fig:sigmoid_comparison}, the recall@1 and recall@5 will peak with the $\epsilon$ value between 5 and 13 depending on the dataset and might drop in performance outside of the range.
The sigmoid value maintain other recalls since their value have already been close to $1.0$. 

\begin{table*}[h]
  \centering
  \begin{tabular}{lllllllll}
    \hline
    \bfseries Loss & 
    \multicolumn{1}{c}{\bfseries Fever} & \multicolumn{1}{c}{\bfseries Wow} & \multicolumn{1}{c}{\bfseries TriviaQA} & \multicolumn{1}{c}{\bfseries Aidayago2} \\
    \cline{2-3}  \cline{4-5} \cline{6-7} \cline{8-9} 
    
    & Recall@5 & Recall@5 & Recall@5 & Recall@5 \\ 
    \hline
    Sigmoid Trick     & \textbf{92.51} & \textbf{82.29} & \textbf{85.47} & \textbf{97.71} \\
    Binary Contrastive     & 79.61 & 81.57 & 84.07 & 96.51 \\
    \hline
  \end{tabular}
  \caption{Comparison between combined sigmoid loss and binary contrastive loss on dev set.}
  \label{table:ablation_table}
\end{table*}

\subsection{Ablation Study}
We performed an ablation study to understand how much our sigmoid loss contributes to the training of ETR. We compare our sigmoid loss to the log contrastive loss by training ETR from FLAN-T5-xl. We save and evaluate the model by comparing ground truth and negatives on the dev set every 10,000 steps, and pick the best-performing model to formally evaluate on the dev set. The result is shown in table \ref{table:ablation_table}.
Across different benchmarks, we found the performance of sigmoid loss consistently better than the log contrastive loss. The experiment result suggests the effectiveness of our sigmoid trick on the task of title reranking.

%% file: sec/6_related_work.tex
\section{Related Works}

\subsection{Document Retrieval and Reranking}

As a solution to improve the performance of retrieval, scholars introduced various reranker solutions. \citet{nogueira2020passage} first proposed using pretrained models to perform document reranking. Later, the paradigm was expanded to Multi-Stage Document Ranking employing BM25 for initial keyword search, monoBERT for document ranking, and duoBERT for document pair-wise scoring \cite{Nogueira:2019}. After that, with the introduction of powerful sequence-to-sequence model T5 \cite{t5_original}, these approaches have also been adopted on T5 for better performances \cite{nogueira2020document}. However, most of these experiments are on MSMARCO dataset \cite{bajaj2018ms}. For KILT benchmark, reranker has also been explored \cite{hofstätter2022fidlight, glass2022re2g}.

\subsection{Training Strategies for Dense Retrieval}

Dense retrieval is a method used in information retrieval that uses vector representations of passages and queries to match their relevance.
Compared with sparse retrieval, it has the advantage of being able to dig into the hidden semantics relations between queries and passages.
There are two mainstream directions for training dense retrievers, i.e., distillation~\cite{ 
izacard2020distilling,
ren2023rocketqav2,
sun2023lead,
zeng2022curriculum,
lin2023prod,
bai2023griprank} and better encoding~\cite{ 
gao2021condenser,
xiao2022retromae,
wu2022contextual,
zhang2023led,
zhuang2022rankt5}.
For distillation techniques, RocketQAv2~\cite{ren2023rocketqav2} presents a joint training approach for DPR and passage
reranking powered by dynamic listwise distillation and hybrid data augmentation.
Besides, PROD~\cite{lin2023prod} gradually improves the student through a teacher progressive and data progressive distillation.
In terms of improving DPR with better encoding methods,
Condenser~\cite{gao2021condenser} proposes a new transformer architecture to establish better encoding.
Meanwhile, CoT-MAE~\cite{wu2022contextual} employs an asymmetric encoder-decoder architecture to encode sentence semantics.

%% file: sec/7_conclusions.tex
\section{Conclusion}

In this work, we propose the Efficient Title Reranker via Broadcasting Query Encoder, a novel efficient reranking technique for knowledge-intensive natural language processing tasks that achieves 20x to 40x speedup over vanilla passage rerankers; in addition, with the assist of our novel Sigmoid Trick, ETR also exceeds in retrieval accuracy and sets new State-of-the-art in four KILT benchmark.

Potential future works include further exploration of ETR in other retrieval tasks. 
Many new large language model application uses tool learning and web search to retrieve information, so one direction of research is to study the effect of title reranking in these tool-learning situations. 
On the other hand, despite our proposed Sigmoid Trick technique was designed for reranker. It also be adopted to wide range of tasks, so another future study can be identifying the tasks that Sigmoid Trick can benefit. 

%% file: sec/8_limitations.tex
\section{Limitations}
Depending on the input query and retrieval corpus, ETR may retrieve incorrect, racial or biased documents.

%% file: acl_latex.bbl
\begin{thebibliography}{44}
\expandafter\ifx\csname natexlab\endcsname\relax\def\natexlab#1{#1}\fi

\bibitem[{Bai et~al.(2023)Bai, Guo, Liu, Yang, Liang, Yan, and Li}]{bai2023griprank}
Jiaqi Bai, Hongcheng Guo, Jiaheng Liu, Jian Yang, Xinnian Liang, Zhao Yan, and Zhoujun Li. 2023.
\newblock \href {http://arxiv.org/abs/2305.18144} {Griprank: Bridging the gap between retrieval and generation via the generative knowledge improved passage ranking}.

\bibitem[{Bajaj et~al.(2018)Bajaj, Campos, Craswell, Deng, Gao, Liu, Majumder, McNamara, Mitra, Nguyen, Rosenberg, Song, Stoica, Tiwary, and Wang}]{bajaj2018ms}
Payal Bajaj, Daniel Campos, Nick Craswell, Li~Deng, Jianfeng Gao, Xiaodong Liu, Rangan Majumder, Andrew McNamara, Bhaskar Mitra, Tri Nguyen, Mir Rosenberg, Xia Song, Alina Stoica, Saurabh Tiwary, and Tong Wang. 2018.
\newblock \href {http://arxiv.org/abs/1611.09268} {Ms marco: A human generated machine reading comprehension dataset}.

\bibitem[{Bevilacqua et~al.(2022)Bevilacqua, Ottaviano, Lewis, Yih, Riedel, and Petroni}]{SEAL}
Michele Bevilacqua, Giuseppe Ottaviano, Patrick Lewis, Scott Yih, Sebastian Riedel, and Fabio Petroni. 2022.
\newblock \href {https://proceedings.neurips.cc/paper_files/paper/2022/file/cd88d62a2063fdaf7ce6f9068fb15dcd-Paper-Conference.pdf} {Autoregressive search engines: Generating substrings as document identifiers}.
\newblock In \emph{Advances in Neural Information Processing Systems}, volume~35, pages 31668--31683. Curran Associates, Inc.

\bibitem[{Brown et~al.(2020)Brown, Mann, Ryder, Subbiah, Kaplan, Dhariwal, Neelakantan, Shyam, Sastry, Askell, Agarwal, Herbert-Voss, Krueger, Henighan, Child, Ramesh, Ziegler, Wu, Winter, Hesse, Chen, Sigler, Litwin, Gray, Chess, Clark, Berner, McCandlish, Radford, Sutskever, and Amodei}]{brown2020language}
Tom~B. Brown, Benjamin Mann, Nick Ryder, Melanie Subbiah, Jared Kaplan, Prafulla Dhariwal, Arvind Neelakantan, Pranav Shyam, Girish Sastry, Amanda Askell, Sandhini Agarwal, Ariel Herbert-Voss, Gretchen Krueger, Tom Henighan, Rewon Child, Aditya Ramesh, Daniel~M. Ziegler, Jeffrey Wu, Clemens Winter, Christopher Hesse, Mark Chen, Eric Sigler, Mateusz Litwin, Scott Gray, Benjamin Chess, Jack Clark, Christopher Berner, Sam McCandlish, Alec Radford, Ilya Sutskever, and Dario Amodei. 2020.
\newblock \href {http://arxiv.org/abs/2005.14165} {Language models are few-shot learners}.

\bibitem[{Cao et~al.(2021)Cao, Izacard, Riedel, and Petroni}]{decao2021autoregressive}
Nicola~De Cao, Gautier Izacard, Sebastian Riedel, and Fabio Petroni. 2021.
\newblock \href {http://arxiv.org/abs/2010.00904} {Autoregressive entity retrieval}.

\bibitem[{Chen et~al.(2022)Chen, Zhang, Guo, Liu, Fan, and Cheng}]{CorpusBrain}
Jiangui Chen, Ruqing Zhang, Jiafeng Guo, Yiqun Liu, Yixing Fan, and Xueqi Cheng. 2022.
\newblock \href {https://doi.org/10.1145/3511808.3557271} {Corpusbrain: Pre-train a generative retrieval model for knowledge-intensive language tasks}.
\newblock In \emph{Proceedings of the 31st ACM International Conference on Information \& Knowledge Management}, CIKM '22, page 191–200, New York, NY, USA. Association for Computing Machinery.

\bibitem[{Child et~al.(2019)Child, Gray, Radford, and Sutskever}]{child2019generating}
Rewon Child, Scott Gray, Alec Radford, and Ilya Sutskever. 2019.
\newblock \href {http://arxiv.org/abs/1904.10509} {Generating long sequences with sparse transformers}.

\bibitem[{Chowdhery et~al.(2022)Chowdhery, Narang, Devlin, Bosma, Mishra, Roberts, Barham, Chung, Sutton, Gehrmann, Schuh, Shi, Tsvyashchenko, Maynez, Rao, Barnes, Tay, Shazeer, Prabhakaran, Reif, Du, Hutchinson, Pope, Bradbury, Austin, Isard, Gur-Ari, Yin, Duke, Levskaya, Ghemawat, Dev, Michalewski, Garcia, Misra, Robinson, Fedus, Zhou, Ippolito, Luan, Lim, Zoph, Spiridonov, Sepassi, Dohan, Agrawal, Omernick, Dai, Pillai, Pellat, Lewkowycz, Moreira, Child, Polozov, Lee, Zhou, Wang, Saeta, Diaz, Firat, Catasta, Wei, Meier-Hellstern, Eck, Dean, Petrov, and Fiedel}]{chowdhery2022palm}
Aakanksha Chowdhery, Sharan Narang, Jacob Devlin, Maarten Bosma, Gaurav Mishra, Adam Roberts, Paul Barham, Hyung~Won Chung, Charles Sutton, Sebastian Gehrmann, Parker Schuh, Kensen Shi, Sasha Tsvyashchenko, Joshua Maynez, Abhishek Rao, Parker Barnes, Yi~Tay, Noam Shazeer, Vinodkumar Prabhakaran, Emily Reif, Nan Du, Ben Hutchinson, Reiner Pope, James Bradbury, Jacob Austin, Michael Isard, Guy Gur-Ari, Pengcheng Yin, Toju Duke, Anselm Levskaya, Sanjay Ghemawat, Sunipa Dev, Henryk Michalewski, Xavier Garcia, Vedant Misra, Kevin Robinson, Liam Fedus, Denny Zhou, Daphne Ippolito, David Luan, Hyeontaek Lim, Barret Zoph, Alexander Spiridonov, Ryan Sepassi, David Dohan, Shivani Agrawal, Mark Omernick, Andrew~M. Dai, Thanumalayan~Sankaranarayana Pillai, Marie Pellat, Aitor Lewkowycz, Erica Moreira, Rewon Child, Oleksandr Polozov, Katherine Lee, Zongwei Zhou, Xuezhi Wang, Brennan Saeta, Mark Diaz, Orhan Firat, Michele Catasta, Jason Wei, Kathy Meier-Hellstern, Douglas Eck, Jeff Dean, Slav Petrov, and Noah Fiedel. 2022.
\newblock \href {http://arxiv.org/abs/2204.02311} {Palm: Scaling language modeling with pathways}.

\bibitem[{Chung et~al.(2022)Chung, Hou, Longpre, Zoph, Tay, Fedus, Li, Wang, Dehghani, Brahma et~al.}]{chung2022scaling}
Hyung~Won Chung, Le~Hou, Shayne Longpre, Barret Zoph, Yi~Tay, William Fedus, Yunxuan Li, Xuezhi Wang, Mostafa Dehghani, Siddhartha Brahma, et~al. 2022.
\newblock Scaling instruction-finetuned language models.
\newblock \emph{arXiv preprint arXiv:2210.11416}.

\bibitem[{{De Cao} et~al.(2021){De Cao}, Izacard, Riedel, and Petroni}]{decao2020autoregressive}
Nicola {De Cao}, Gautier Izacard, Sebastian Riedel, and Fabio Petroni. 2021.
\newblock \href {https://openreview.net/forum?id=5k8F6UU39V} {Autoregressive entity retrieval}.
\newblock In \emph{International Conference on Learning Representations}.

\bibitem[{Devlin et~al.(2019)Devlin, Chang, Lee, and Toutanova}]{devlin-etal-2019-bert}
Jacob Devlin, Ming-Wei Chang, Kenton Lee, and Kristina Toutanova. 2019.
\newblock \href {https://doi.org/10.18653/v1/N19-1423} {{BERT}: Pre-training of deep bidirectional transformers for language understanding}.
\newblock In \emph{Proceedings of the 2019 Conference of the North {A}merican Chapter of the Association for Computational Linguistics: Human Language Technologies, Volume 1 (Long and Short Papers)}, pages 4171--4186, Minneapolis, Minnesota. Association for Computational Linguistics.

\bibitem[{Dinan et~al.(2018)Dinan, Roller, Shuster, Fan, Auli, and Weston}]{dinan2018wizard}
Emily Dinan, Stephen Roller, Kurt Shuster, Angela Fan, Michael Auli, and Jason Weston. 2018.
\newblock Wizard of wikipedia: Knowledge-powered conversational agents.
\newblock \emph{arXiv preprint arXiv:1811.01241}.

\bibitem[{Fu et~al.(2023)Fu, Dao, Saab, Thomas, Rudra, and Ré}]{fu2023hungry}
Daniel~Y. Fu, Tri Dao, Khaled~K. Saab, Armin~W. Thomas, Atri Rudra, and Christopher Ré. 2023.
\newblock \href {http://arxiv.org/abs/2212.14052} {Hungry hungry hippos: Towards language modeling with state space models}.

\bibitem[{Gao and Callan(2021)}]{gao2021condenser}
Luyu Gao and Jamie Callan. 2021.
\newblock \href {http://arxiv.org/abs/2104.08253} {Condenser: a pre-training architecture for dense retrieval}.

\bibitem[{Glass et~al.(2022)Glass, Rossiello, Chowdhury, Naik, Cai, and Gliozzo}]{glass2022re2g}
Michael Glass, Gaetano Rossiello, Md~Faisal~Mahbub Chowdhury, Ankita~Rajaram Naik, Pengshan Cai, and Alfio Gliozzo. 2022.
\newblock \href {http://arxiv.org/abs/2207.06300} {Re2g: Retrieve, rerank, generate}.

\bibitem[{Gu and Dao(2023)}]{gu2023mamba}
Albert Gu and Tri Dao. 2023.
\newblock \href {http://arxiv.org/abs/2312.00752} {Mamba: Linear-time sequence modeling with selective state spaces}.

\bibitem[{Hofstätter et~al.(2022)Hofstätter, Chen, Raman, and Zamani}]{hofstätter2022fidlight}
Sebastian Hofstätter, Jiecao Chen, Karthik Raman, and Hamed Zamani. 2022.
\newblock \href {http://arxiv.org/abs/2209.14290} {Fid-light: Efficient and effective retrieval-augmented text generation}.

\bibitem[{Izacard and Grave(2020)}]{izacard2020distilling}
Gautier Izacard and Edouard Grave. 2020.
\newblock \href {http://arxiv.org/abs/2012.04584} {Distilling knowledge from reader to retriever for question answering}.

\bibitem[{Joshi et~al.(2017)Joshi, Choi, Weld, and Zettlemoyer}]{joshi2017triviaqa}
Mandar Joshi, Eunsol Choi, Daniel~S Weld, and Luke Zettlemoyer. 2017.
\newblock Triviaqa: A large scale distantly supervised challenge dataset for reading comprehension.
\newblock \emph{arXiv preprint arXiv:1705.03551}.

\bibitem[{Karpukhin et~al.(2020)Karpukhin, Oğuz, Min, Lewis, Wu, Edunov, Chen, and tau Yih}]{karpukhin2020dense}
Vladimir Karpukhin, Barlas Oğuz, Sewon Min, Patrick Lewis, Ledell Wu, Sergey Edunov, Danqi Chen, and Wen tau Yih. 2020.
\newblock \href {http://arxiv.org/abs/2004.04906} {Dense passage retrieval for open-domain question answering}.

\bibitem[{Lewis et~al.(2020)Lewis, Perez, Piktus, Petroni, Karpukhin, Goyal, K{\"u}ttler, Lewis, Yih, Rockt{\"a}schel et~al.}]{lewis2020retrieval}
Patrick Lewis, Ethan Perez, Aleksandra Piktus, Fabio Petroni, Vladimir Karpukhin, Naman Goyal, Heinrich K{\"u}ttler, Mike Lewis, Wen-tau Yih, Tim Rockt{\"a}schel, et~al. 2020.
\newblock Retrieval-augmented generation for knowledge-intensive nlp tasks.
\newblock \emph{Advances in Neural Information Processing Systems}, 33:9459--9474.

\bibitem[{Lewis et~al.(2021)Lewis, Perez, Piktus, Petroni, Karpukhin, Goyal, Küttler, Lewis, tau Yih, Rocktäschel, Riedel, and Kiela}]{lewis2021retrievalaugmented}
Patrick Lewis, Ethan Perez, Aleksandra Piktus, Fabio Petroni, Vladimir Karpukhin, Naman Goyal, Heinrich Küttler, Mike Lewis, Wen tau Yih, Tim Rocktäschel, Sebastian Riedel, and Douwe Kiela. 2021.
\newblock \href {http://arxiv.org/abs/2005.11401} {Retrieval-augmented generation for knowledge-intensive nlp tasks}.

\bibitem[{Lin et~al.(2023{\natexlab{a}})Lin, Gong, Liu, Zhang, Lin, Dong, Jiao, Lu, Jiang, Majumder, and Duan}]{10.1145/3543507.3583421}
Zhenghao Lin, Yeyun Gong, Xiao Liu, Hang Zhang, Chen Lin, Anlei Dong, Jian Jiao, Jingwen Lu, Daxin Jiang, Rangan Majumder, and Nan Duan. 2023{\natexlab{a}}.
\newblock \href {https://doi.org/10.1145/3543507.3583421} {Prod: Progressive distillation for dense retrieval}.
\newblock In \emph{Proceedings of the ACM Web Conference 2023}, WWW '23, page 3299–3308, New York, NY, USA. Association for Computing Machinery.

\bibitem[{Lin et~al.(2023{\natexlab{b}})Lin, Gong, Liu, Zhang, Lin, Dong, Jiao, Lu, Jiang, Majumder, and Duan}]{lin2023prod}
Zhenghao Lin, Yeyun Gong, Xiao Liu, Hang Zhang, Chen Lin, Anlei Dong, Jian Jiao, Jingwen Lu, Daxin Jiang, Rangan Majumder, and Nan Duan. 2023{\natexlab{b}}.
\newblock \href {http://arxiv.org/abs/2209.13335} {Prod: Progressive distillation for dense retrieval}.

\bibitem[{Maillard et~al.(2021)Maillard, Karpukhin, Petroni, Yih, Oguz, Stoyanov, and Ghosh}]{MultiDPR}
Jean Maillard, Vladimir Karpukhin, Fabio Petroni, Wen-tau Yih, Barlas Oguz, Veselin Stoyanov, and Gargi Ghosh. 2021.
\newblock \href {https://doi.org/10.18653/v1/2021.acl-long.89} {Multi-task retrieval for knowledge-intensive tasks}.
\newblock In \emph{Proceedings of the 59th Annual Meeting of the Association for Computational Linguistics and the 11th International Joint Conference on Natural Language Processing (Volume 1: Long Papers)}, pages 1098--1111, Online. Association for Computational Linguistics.

\bibitem[{Nogueira and Cho(2020)}]{nogueira2020passage}
Rodrigo Nogueira and Kyunghyun Cho. 2020.
\newblock \href {http://arxiv.org/abs/1901.04085} {Passage re-ranking with bert}.

\bibitem[{Nogueira et~al.(2020)Nogueira, Jiang, and Lin}]{nogueira2020document}
Rodrigo Nogueira, Zhiying Jiang, and Jimmy Lin. 2020.
\newblock \href {http://arxiv.org/abs/2003.06713} {Document ranking with a pretrained sequence-to-sequence model}.

\bibitem[{Nogueira et~al.(2019{\natexlab{a}})Nogueira, Yang, Cho, and Lin}]{nogueira2019multistage}
Rodrigo Nogueira, Wei Yang, Kyunghyun Cho, and Jimmy Lin. 2019{\natexlab{a}}.
\newblock \href {http://arxiv.org/abs/1910.14424} {Multi-stage document ranking with bert}.

\bibitem[{Nogueira et~al.(2019{\natexlab{b}})Nogueira, Yang, Cho, and Lin}]{Nogueira:2019}
Rodrigo Nogueira, Wei Yang, Kyunghyun Cho, and Jimmy Lin. 2019{\natexlab{b}}.
\newblock \href {https://arxiv.org/abs/1910.14424} {Multi-stage document ranking with bert}.
\newblock \emph{arXiv preprint arXiv:1910.14424}.
\newblock Submitted on 31 Oct 2019.

\bibitem[{OpenAI(2023)}]{openai2023gpt4}
OpenAI. 2023.
\newblock \href {http://arxiv.org/abs/2303.08774} {Gpt-4 technical report}.

\bibitem[{Petroni et~al.(2021)Petroni, Piktus, Fan, Lewis, Yazdani, Cao, Thorne, Jernite, Karpukhin, Maillard, Plachouras, Rocktäschel, and Riedel}]{petroni2021kilt}
Fabio Petroni, Aleksandra Piktus, Angela Fan, Patrick Lewis, Majid Yazdani, Nicola~De Cao, James Thorne, Yacine Jernite, Vladimir Karpukhin, Jean Maillard, Vassilis Plachouras, Tim Rocktäschel, and Sebastian Riedel. 2021.
\newblock \href {http://arxiv.org/abs/2009.02252} {Kilt: a benchmark for knowledge intensive language tasks}.

\bibitem[{Raffel et~al.(2020)Raffel, Shazeer, Roberts, Lee, Narang, Matena, Zhou, Li, and Liu}]{t5_original}
Colin Raffel, Noam Shazeer, Adam Roberts, Katherine Lee, Sharan Narang, Michael Matena, Yanqi Zhou, Wei Li, and Peter~J. Liu. 2020.
\newblock Exploring the limits of transfer learning with a unified text-to-text transformer.
\newblock \emph{J. Mach. Learn. Res.}, 21(1).

\bibitem[{Ren et~al.(2021)Ren, Qu, Liu, Zhao, She, Wu, Wang, and Wen}]{ren-etal-2021-rocketqav2}
Ruiyang Ren, Yingqi Qu, Jing Liu, Wayne~Xin Zhao, QiaoQiao She, Hua Wu, Haifeng Wang, and Ji-Rong Wen. 2021.
\newblock \href {https://doi.org/10.18653/v1/2021.emnlp-main.224} {{R}ocket{QA}v2: A joint training method for dense passage retrieval and passage re-ranking}.
\newblock In \emph{Proceedings of the 2021 Conference on Empirical Methods in Natural Language Processing}, pages 2825--2835, Online and Punta Cana, Dominican Republic. Association for Computational Linguistics.

\bibitem[{Ren et~al.(2023)Ren, Qu, Liu, Zhao, She, Wu, Wang, and Wen}]{ren2023rocketqav2}
Ruiyang Ren, Yingqi Qu, Jing Liu, Wayne~Xin Zhao, Qiaoqiao She, Hua Wu, Haifeng Wang, and Ji-Rong Wen. 2023.
\newblock \href {http://arxiv.org/abs/2110.07367} {Rocketqav2: A joint training method for dense passage retrieval and passage re-ranking}.

\bibitem[{Sun et~al.(2023{\natexlab{a}})Sun, Liu, Gong, Dong, Lu, Zhang, Yang, Majumder, and Duan}]{sun2023lead}
Hao Sun, Xiao Liu, Yeyun Gong, Anlei Dong, Jingwen Lu, Yan Zhang, Linjun Yang, Rangan Majumder, and Nan Duan. 2023{\natexlab{a}}.
\newblock \href {http://arxiv.org/abs/2212.05225} {Lead: Liberal feature-based distillation for dense retrieval}.

\bibitem[{Sun et~al.(2023{\natexlab{b}})Sun, Yan, Ma, Wang, Ren, Chen, Yin, and Ren}]{sun-etal-2023-chatgpt}
Weiwei Sun, Lingyong Yan, Xinyu Ma, Shuaiqiang Wang, Pengjie Ren, Zhumin Chen, Dawei Yin, and Zhaochun Ren. 2023{\natexlab{b}}.
\newblock \href {https://doi.org/10.18653/v1/2023.emnlp-main.923} {Is {C}hat{GPT} good at search? investigating large language models as re-ranking agents}.
\newblock In \emph{Proceedings of the 2023 Conference on Empirical Methods in Natural Language Processing}, pages 14918--14937, Singapore. Association for Computational Linguistics.

\bibitem[{Thorne et~al.(2018)Thorne, Vlachos, Christodoulopoulos, and Mittal}]{thorne2018fever}
James Thorne, Andreas Vlachos, Christos Christodoulopoulos, and Arpit Mittal. 2018.
\newblock Fever: a large-scale dataset for fact extraction and verification.
\newblock \emph{arXiv preprint arXiv:1803.05355}.

\bibitem[{Wang et~al.(2023)Wang, Liu, Yue, Tang, Zhang, Jiayang, Yao, Gao, Hu, Qi, Wang, Yang, Wang, Xie, Zhang, and Zhang}]{wang2023survey}
Cunxiang Wang, Xiaoze Liu, Yuanhao Yue, Xiangru Tang, Tianhang Zhang, Cheng Jiayang, Yunzhi Yao, Wenyang Gao, Xuming Hu, Zehan Qi, Yidong Wang, Linyi Yang, Jindong Wang, Xing Xie, Zheng Zhang, and Yue Zhang. 2023.
\newblock \href {http://arxiv.org/abs/2310.07521} {Survey on factuality in large language models: Knowledge, retrieval and domain-specificity}.

\bibitem[{Wu et~al.(2022)Wu, Ma, Lin, Lin, Wang, and Hu}]{wu2022contextual}
Xing Wu, Guangyuan Ma, Meng Lin, Zijia Lin, Zhongyuan Wang, and Songlin Hu. 2022.
\newblock \href {http://arxiv.org/abs/2208.07670} {Contextual masked auto-encoder for dense passage retrieval}.

\bibitem[{Xiao et~al.(2022)Xiao, Liu, Shao, and Cao}]{xiao2022retromae}
Shitao Xiao, Zheng Liu, Yingxia Shao, and Zhao Cao. 2022.
\newblock \href {http://arxiv.org/abs/2205.12035} {Retromae: Pre-training retrieval-oriented language models via masked auto-encoder}.

\bibitem[{Zeng et~al.(2022)Zeng, Zamani, and Vinay}]{zeng2022curriculum}
Hansi Zeng, Hamed Zamani, and Vishwa Vinay. 2022.
\newblock \href {http://arxiv.org/abs/2204.13679} {Curriculum learning for dense retrieval distillation}.

\bibitem[{Zhang et~al.(2023{\natexlab{a}})Zhang, Tao, Shen, Xu, Geng, Jiao, and Jiang}]{zhang2023led}
Kai Zhang, Chongyang Tao, Tao Shen, Can Xu, Xiubo Geng, Binxing Jiao, and Daxin Jiang. 2023{\natexlab{a}}.
\newblock \href {http://arxiv.org/abs/2208.13661} {Led: Lexicon-enlightened dense retriever for large-scale retrieval}.

\bibitem[{Zhang et~al.(2023{\natexlab{b}})Zhang, Li, Cui, Cai, Liu, Fu, Huang, Zhao, Zhang, Chen et~al.}]{zhang2023siren}
Yue Zhang, Yafu Li, Leyang Cui, Deng Cai, Lemao Liu, Tingchen Fu, Xinting Huang, Enbo Zhao, Yu~Zhang, Yulong Chen, et~al. 2023{\natexlab{b}}.
\newblock Siren's song in the ai ocean: A survey on hallucination in large language models.
\newblock \emph{arXiv preprint arXiv:2309.01219}.

\bibitem[{Zhuang et~al.(2022)Zhuang, Qin, Jagerman, Hui, Ma, Lu, Ni, Wang, and Bendersky}]{zhuang2022rankt5}
Honglei Zhuang, Zhen Qin, Rolf Jagerman, Kai Hui, Ji~Ma, Jing Lu, Jianmo Ni, Xuanhui Wang, and Michael Bendersky. 2022.
\newblock \href {http://arxiv.org/abs/2210.10634} {Rankt5: Fine-tuning t5 for text ranking with ranking losses}.

\end{thebibliography}
